\documentclass[aip,rsi,reprint]{revtex4-1}
\usepackage{graphicx}
\usepackage{dcolumn}
\usepackage{bm}

\begin{document}

\title{Single exposure 3D imaging of dusty plasma clusters}

\author{Peter Hartmann}
\email{hartmann.peter@wigner.mta.hu}
\affiliation{Institute for Solid State Physics and Optics, Wigner Research Centre for Physics, Hungarian Academy of Sciences, P.O.B. 49, H-1525 Budapest, Hungary}
\affiliation{Center for Astrophysics, Space Physics and Engineering Research (CASPER), One Bear Place 97310, Baylor University, Waco, TX 76798, USA}

\author{Istv\'an Donk\'o}
\affiliation{Illy\'es Gyula High School, 2040 Buda\"ors, Szabads\'ag \'ut 162., Hungary}

\author{Zolt\'an Donk\'o}
\affiliation{Institute for Solid State Physics and Optics, Wigner Research Centre for Physics, Hungarian Academy of Sciences, P.O.B. 49, H-1525 Budapest, Hungary}

\date{\today}

\begin{abstract}

We have worked out the details of a single camera, single exposure method to perform three-dimensional imaging of a finite particle cluster. The procedure is based on the plenoptic imaging principle and utilizes a commercial Lytro light field still camera. We demonstrate the capabilities of our technique on a single layer particle cluster in a dusty plasma, where the camera is aligned inclined at a small angle to the particle layer. The reconstruction of the third coordinate (depth) is found to be accurate and even shadowing particles can be identified.

\end{abstract}

\pacs{52.27.Lw, 07.05.Pj, 42.30.Wb}

\maketitle 

\section{Introduction}

Three dimensional imaging is in general a very challenging task. With the rapid development of digital imaging and fast data processing capabilities, this field experienced a boom in the recent years. The field of dusty plasma research is of no exception. Since the first parallel reports on the observation of ``plasma crystals'' \cite{pk1,pk2,pk3}, which are usually situated in a horizontal plane, the main diagnostic tool in this field has been video microscopy combined with particle tracking velocimetry (PTV). The need for accurate three-dimensional imaging was recognized soon after, when microgravity experiments started in space \cite{Fortov,Nefedov}, on parabolic flights \cite{parabolic}, as well as with the discovery of Coulomb balls \cite{Cball} and other ``vertically extended'' systems in the laboratory \cite{WacoBox}. With 3D imaging being a generally unsolved issue, there is no straightforward ``best'' solution to record and track particle positions in these systems. Four different methods, of which the development was partially motivated by this field, have been successfully applied to analyze dust particle configurations. These are: the slicing method \cite{slicing}, stereoscopy \cite{stereo}, the color gradient method \cite{gradient}, and digital holography \cite{holography}. All of these techniques, briefly reviewed in the next section, have their pros and cons. 

The aim of this work is to elaborate the details and to demonstrate the applicability of a new method utilizing a single camera, where all three spatial coordinates can be reconstructed from a single short exposure of a dust particle cloud. The optical principle is known as the ``plenoptic technique'', proposed already in 1908 by Gabriel Lippmann \cite{Lippmann} (who called it ``integral photography''). It took, however over a century for his invention to evolve from an idea to a commercial product. In the year 2011 two companies introduced their solutions naming them ``light field'' cameras. The {\it Lytro} is a low cost, point-and-shoot type still camera, while the {\it Raytrix} is an industrial grade video solution with up to 180 frames per second at megapixel image resolution. Light field cameras provide the possibility to refocus the image after exposure by virtually displacing the image plane \cite{Ng2005}. 

Following a short introduction in Section \ref{sec:meth} to the existing methods mentioned above, in Section \ref{sec:DP} we introduce our dusty plasma experiment. This is followed by the detailed description of our image processing and particle tracking method, in Section \ref{sec:proc}.

\section{Existing methods}\label{sec:meth}

\subsection{Slicing method} 
The slicing method is in principle a simple 2D method, where the target is illuminated by a thin laser sheet and the light scattered from the particles is captured by a digital camera positioned perpendicular to that layer. By moving the illumination together with the camera back and forth, a series of 2D slices are recorded. The reconstruction of the depth is accomplished simply by associating the slice number to the depth position of the illumination \cite{slicing}. 

Advantage:
\begin{itemize}
\item Low sensitivity to general lens properties like depth of field and perspective.
\end{itemize}

Disadvantages:
\begin{itemize}
\item Slow, as many exposures are needed for a single scan.
\item Low depth resolution.
\item Useful on static or very slowly evolving systems only, as every slice belongs to a different instance in time.
\end{itemize}

\subsection{Stereoscopy} 
Stereoscopy is the most intuitive and most popular technique, as it resembles the way of 3D vision as existing in nature. The principle is simple: pointing two or more cameras to the same observation volume from different directions allows, in principle, perfect 3D reconstruction \cite{stereo}.

Advantage:
\begin{itemize}
\item High frame-rates are possible (in typical applications up to 150 fps), limited by the individual cameras, which have to operate synchronized.
\end{itemize}

Disadvantages:
\begin{itemize}
\item Expensive, as the cost increases linearly with the number of cameras used.
\item Low depth of field or low sensitivity due to low aperture (high f-number) used.
\item Problems with shadowing particles
\item Perspective problems. Can be solved by using telecentric lenses, but those increase the installation expenses significantly.
\end{itemize}

\subsection{Color gradient method} 
The color gradient method uses two (or more) cameras, which observe the volume of interest from the same direction but using complementing color filters. The illumination of the particle ensemble is performed using two colors with linear and opposite intensity gradients. For example the intensity of the {\it red} light decreases with increasing distance from the camera, while the {\it blue} intensity increases in the same direction. This way the depth information is simply given by the ratio of the apparent {\it red} and {\it blue} color intensities measured on each individual particle \cite{gradient}. 

Advantages:
\begin{itemize}
\item Simple mathematics needed for the reconstruction, which is less sensitive to lens and perspective problems compared to stereoscopy.
\item Can be as fast as stereoscopy
\end{itemize}

Disadvantages:
\begin{itemize}
\item The use of laser sources for illumination seems to be straightforward, but due to their {\it near perfect} beam properties the strong angular dependence of the scattered light intensity, as described by the Mie scattering model, can result in misleading conclusions.
\item High dynamic range (12 or 16 bit in intensity) needed to achieve a depth resolution comparable to the horizontal and vertical resolution.
\item Problems with the limited depth of field.
\item Problems with shadowing particles.
\end{itemize}

\subsection{Digital in-line holography} 
The digital in-line holography method represents a completely different approach from any conventional imaging techniques. The volume of interest is illuminated by a high quality, wide, single mode laser beam. As the laser light is scattered on the levitating particles, a small fraction of it forms interference rings, which fall on a digital image sensor. During the analysis of the images the center position of the rings give the horizontal and vertical coordinates of a particle, while the depth has to be computed from the interference ring structure \cite{holography}.

Advantage:
\begin{itemize}
\item No lens distortions, as no optics at all is involved.
\end{itemize}

Disadvantages:
\begin{itemize}
\item Numerically very demanding reconstruction.
\item Very high dynamic range (16 bit or more in intensity) needed to capture the faint interference ring structure on the background of the direct laser beam.
\item Very large area and high resolution detector needed to capture as much of the interference rings as possible.
\item Slow: demonstrated so far only on static dust clusters with low particle number and large particle sizes.
\end{itemize}

As one can see from the introduction of the available techniques, there is no general solution. The optimal choice strongly depends on the specific properties of the system under investigation and the quantities of interest.

\section{Dusty plasma experiment} \label{sec:DP} 

\begin{figure}[htb]
\includegraphics[width=8cm]{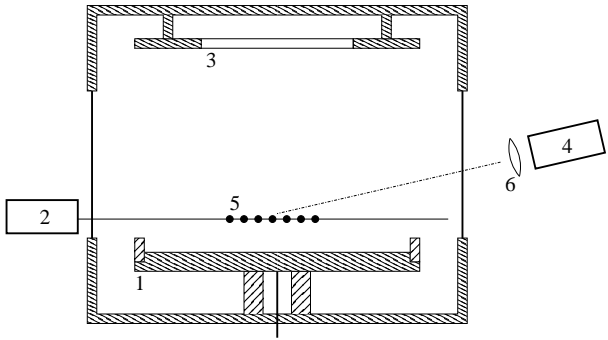}
\caption{\label{fig:exp}
Schematics of the dusty plasma experiment. 1: powered electrode, 2: illuminating laser (200~mW @ 532~nm), 3: grounded electrode, 4: {\it Lytro} camera, 5: dust particle cluster, 6: $f=174$~mm convex lens to shorten the working distance.}
\end{figure}

Our dusty plasma experiments are carried out in a custom designed vacuum chamber with an inner diameter of 25~cm and a height of 18~cm. The lower, powered, 17~cm diameter, flat, horizontal, stainless steel electrode faces the upper, ring shaped, grounded aluminum electrode, which has an inner diameter of 15~cm and is positioned at a height of 13~cm. The experiments are performed in an argon gas discharge at a pressure $p = 1.1\pm0.05$~Pa, at a steady gas flow of a few times 0.01~sccm, with 13.56~MHz radio frequency excitation of ca. 10~W power. Melamine-formaldehyde micro-spheres with a diameter $d = 9.16\pm0.09~\mu$m are used. For illumination of the particle layer we use a 200~mW, 532~nm (green) laser, the light of which is expanded and enters the chamber through a side window. Although the present work targets 3D imaging, to provide a reliable reference we have chosen to test our technique on a medium size (about 14~mm diameter) single layer dust cluster consisting of about 60 particles. The light field camera captures its images from the side with ca. $13^{\circ}$ tilt angle to the dust particle layer, as shown in figure \ref{fig:exp}. This configuration represents a test case, which makes the verification of the depth measurement possible.   

\begin{figure}[htb]
\includegraphics[width=7cm]{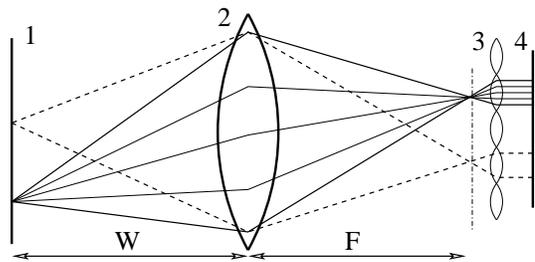}
\caption{\label{fig:opt}
Schematics of the optical configuration of the light field camera (not to scale). 1: world plane, 2: objective lens, 3: micro-lens array, 4: CMOS sensor array, $W$: working distance, $F$: distance between the objective lens and the image plane. Light rays from different world plane points fall on different micro-lenses (rays with solid vs. dashed lines), while rays originating from the same points of the world plane and passing the objective lens at different points (e.g. the rays shown by solid lines) fall on different sensor pixels behind a given micro-lens.}
\end{figure}

The {\it Lytro} camera is, in fact, not much different from a usual compact digital camera. It has a $3280\times 3280$ CMOS sensor with a pixel size of $1.4~\mu$m, a (RG:GB) Bayer color filter matrix, and a 12 bit analog to digital converter, as well as a zoom lens with a constant $f/2$ aperture. The most important difference is, that in front of the CMOS sensor, at about $25~\mu$m distance an array of micro-lenses is mounted. The micro-lenses of about $14~\mu$m in diameter form a triangular lattice. This design enables to compute the light field function $L_F(s,t,u,v)$, which gives the light intensity arriving at the detector coordinates $(s,t)$ from the position $(u,v)$ of the objective lens, as illustrated in figure \ref{fig:opt}. In other words, each micro-lens projects the objective lens onto the set of detector pixels situated behind it, thus each sensor pixel measures the light intensity that has entered the camera through a specific point $(u,v)$ of the objective lens and impacted a specific micro-lens with coordinates $(s,t)$.

\section{Image processing and particle detection}  \label{sec:proc} 

Light field cameras use wide aperture lenses, thus have a narrow depth of field, much shorter than the diameter of the dust particle cloud in our dusty plasma experiment. As a consequence, particles within the depth of field appear as bright points in the image, while particles situated closer to, or further away from the camera show up as faint blurred blobs. The actual intensity profile produced by these out-of-focus particles (called {\it bokeh} in photography) depends strongly on the properties of the objective lens. To obtain the three dimensional coordinates of each particle we perform the following logical steps:
\begin{enumerate}
\item Compute refocused images representing different depth layers from the light field function (the core concept of light field photography \cite{Ng2005}). This way we obtain a series of images with working distances $W$ scanning through the dust particle cloud.  
\item Measure the apparent brightness $B$, and central coordinates $(x,y)$ of all observable particle projections on every image.
\item Interpolate $B$ versus $W$ and find $W_i$ which maximizes $B_i$, where the index $i$ labels the particles within the cloud. Once found, $W_i$ is equal to the $z$ (depth) coordinate of particle $i$, while the $(x,y)_i$ (world plane) coordinates are found from simple interpolation of the measured values to $W_i$.
\end{enumerate}

The first step of implementing the refocusing procedure is to construct a look-up-table (LOT) that correctly associates the four dimensional coordinates $(s,t,u,v)$ to each and every sensor pixel of interest. The LOT is unique to each camera and is independent of the exposure, thus has to be constructed only once. Here we recall, that we use green (532~nm) illumination of our dust cloud, thus in the following we process only pixels behind the green color filters of the Bayer matrix, which is exactly half of the total sensor pixels. To construct the LOT, important calibration information is needed, which can be found in the header section of the raw image files, downloadable from the camera (like angular misalignment and offset of the micro-lens array, pixel and lens pitch values, etc.). Using the LOT, the light field function $L_F(s,t,u,v)$ belonging to an exposure can be constructed.

With the light field function in hand the computation of the primary 2D image (as exposed) is possible based on the numerical evaluation of the integral projection expression
\begin{equation}\label{eq:1}
E_F(s,t)=\frac{1}{F^2}\int\int L_F(s,t,u,v) \cos^4\phi ~{\rm d}u {\rm d}v,
\end{equation}
where $E_F(s,t)$ is the monochromatic 2D image, $F$ is the distance between the objective lens and the sensor plane, $\phi$ is the angle between the incident ray and the optical axis and is purely a geometrical factor independent of the actual exposure. The integration runs over the open aperture of the objective lens \cite{Ng2005}.

Refocusing is introduced by virtually shifting the image plane distance $F$ to $F^\prime = \alpha F$. In this case the light field function is transformed as $L_{F^\prime}(s,t,u,v) = L_{\alpha F}(s,t,u,v) = L_F(u+(s-u)/\alpha,v+(t-v)/\alpha,u,v)$, and the 2D projection formula changes to \cite{Ng2005}:
\begin{eqnarray}
E_{\alpha F}(s,t) &=&\frac{1}{\alpha^2 F^2} \int\int \cos^4\phi \\
&& L_F\left[u+(s-u)/\alpha,v+(t-v)/\alpha,u,v\right] ~{\rm d}u {\rm d}v.\nonumber 
\end{eqnarray} 

The evaluation of these integrals is performed numerically, discretising them using the LOT pre-constructed for the particular camera. As each micro-lens projects the main lens onto the sensor pixels behind it, the discrete $(u,v)$ main lens coordinates are obtained by 
\begin{eqnarray}
u&=&\beta \Delta x + L_x \\
v&=&\beta \Delta y + L_y, \nonumber
\end{eqnarray}
where $\beta$ is a magnification factor (approximately the ratio of $F$ and the focal length of the micro-lenses) taken from the calibration information of the camera, $\Delta x$ and $\Delta y$ are the relative coordinates of the sensor pixels to the centre of the corresponding micro-lens, $L_x$ and $L_y$ are the relative coordinates of the centers of the corresponding micro-lens to the optical axes of the camera.

To optimize the computations the discrete values of $(s,t)$ are chosen corresponding to the centers of the micro-lenses, which form a triangular lattice. The virtually refocused 2D images are results of barycentic interpolations of the computed $E_{\alpha F}(s,t)$ intensity maps for a series of $\alpha$ parameters.

\begin{figure}[htb]
\setlength{\fboxsep}{0pt}
\setlength{\fboxrule}{1pt}
\fbox{\includegraphics[width=7cm]{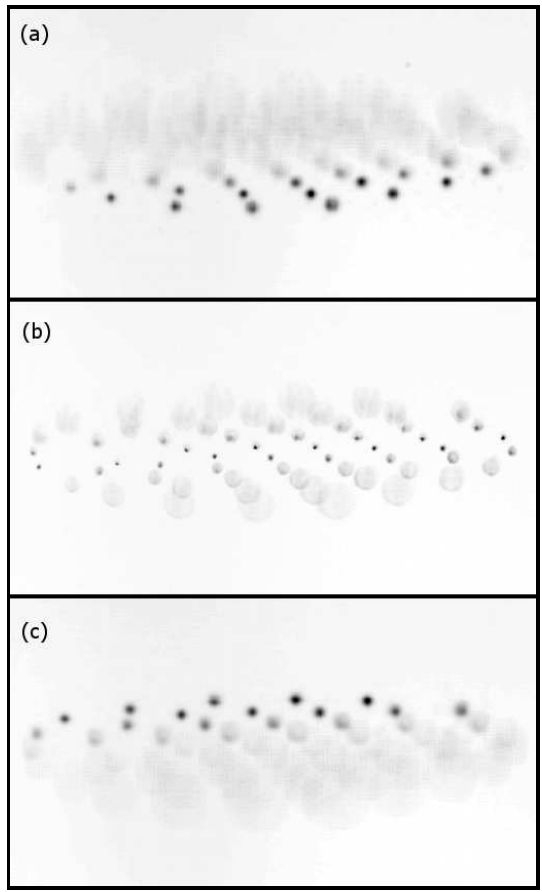}}
\caption{\label{fig:slices}
Inverted and enhanced contrast images computed for $\alpha = 0.987$, 1.0, and 1.013 (from top to bottom) from a single exposure.}
\end{figure}

For our first benchmarking experiment we have computed 40 virtually refocused images from a single exposure. Figure \ref{fig:slices} shows three selected cases to illustrate the capabilities of our image processing algorithm. The centre image is the one seen on the raw image before any refocusing. In this image only particles situated at ``medium'' distances show up sharply as the camera was focused at the centre of the dust particle cloud.

\begin{figure}[htb]
\setlength{\fboxsep}{0pt}
\setlength{\fboxrule}{1pt}
\fbox{\includegraphics[width=7cm]{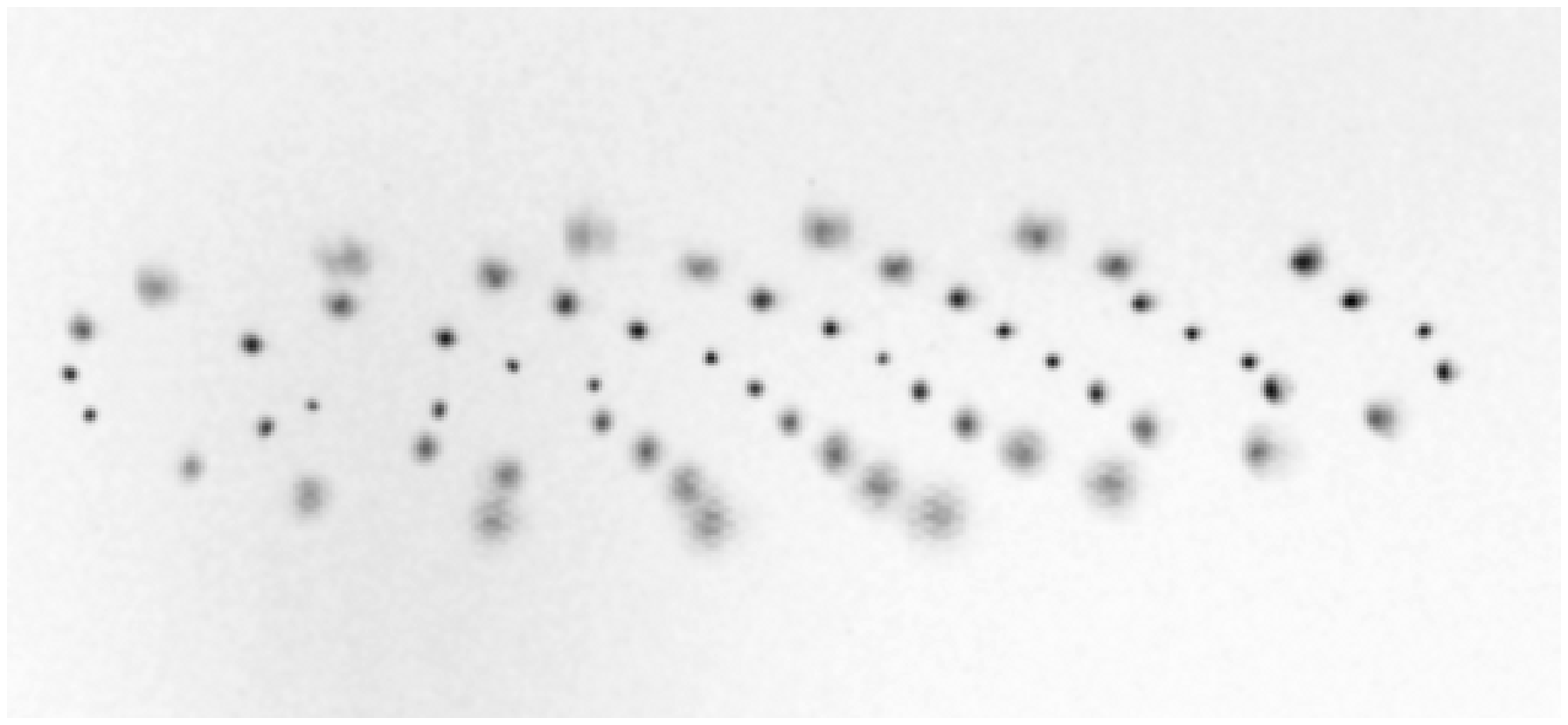}}
\caption{\label{fig:sub}
Sub-aperture (enhanced field of depth) image.}
\end{figure}

Before we preform the particle detection in each image, we make benefit from another possibility offered by the light field technique, namely the ``digital stepping-down'' of the image simply by constricting the integration in eq. (\ref{eq:1}) to a small part of the main lens. The sub-aperture image computed this way has an enhanced field of depth for the price of higher noise level, which can be reduced by applying a Gaussian blur filter, as shown in figure \ref{fig:sub}. The multiplication of the refocused images with the sub-aperture image significantly enhances the apparent brightness of the particles in the vicinity of the field of depth relative to the unfocused ones.

\begin{figure}[htb]
\includegraphics[width=8cm]{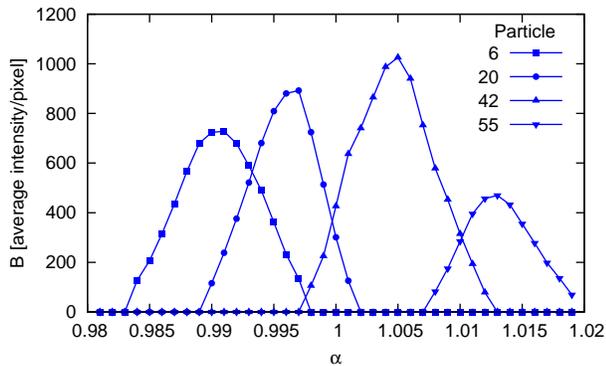}
\caption{\label{fig:bright}
Examples of $B_i(\alpha)$ brightness functions for four representative particles.}
\end{figure}

Particle detection is performed in these multiplied images applying the widely used momentum method \cite{moment}. Besides the $(x,y)$ coordinates of the particles identified in each image, the apparent brightness (defined as the average intensity per pixel) of each particle is recorded as well. After identifying corresponding particles on subsequent images, the brightness function $B_i(\alpha)$ for each particle can be constructed. A few examples are shown in figure \ref{fig:bright}.

\begin{figure}[htb]
\includegraphics[width=6.5cm]{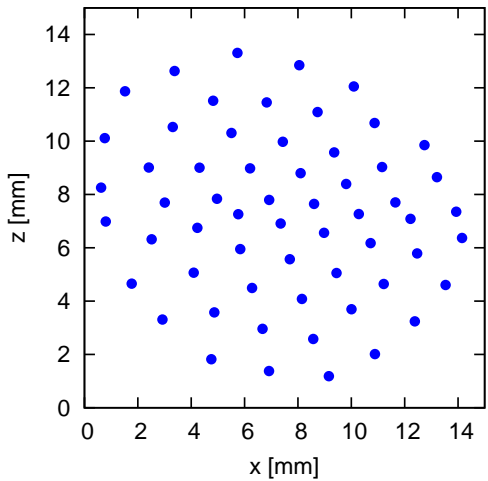}
\caption{\label{fig:part}
Top view of the dust particle cloud projected from the full 3D coordinate set.}
\end{figure}

\begin{figure}[htb]
\includegraphics[width=8cm]{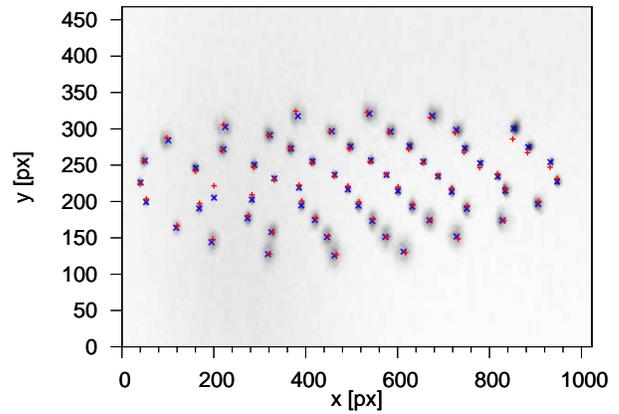}
\caption{\label{fig:bench}
Overlay of the sub-aperture image, apparent $(x,y)$ coordinates (blue crosses), and the tilted 3D particle coordinate projections (red crosses).}
\end{figure}

The position of the maxima of the $B_i(\alpha)$ brightness functions determine the $\alpha_i$ parameters, which represent the particles' depth coordinate relative to the original working distance of the objective lens. After calibrating the apparent pixel sizes on the images to the physical measures of the dust particle cloud, the absolute $z$ (depth) coordinate can be determined. Figure \ref{fig:part} shows the top view of the depth reconstructed dust particle cloud, while figure \ref{fig:bench} shows and an overlay of the sub-aperture image and the projected particle coordinates to demonstrate the accuracy of our algorithm.

The quantitative comparison of the apparent 2D and the projected 3D coordinates shows, that the depth measurement has an uncertainty (standard deviation) of ca. 7\% of the apparent inter-particle distance, which is 4~times higher than that of the 2D $(x,y)$ coordinates, which can be assumed to be 1~image pixel. This accuracy is comparable to that of other techniques, and further improvements are expected with the fine-tuning of our algorithm, and further advances of the light-field technique.

Furthermore, this technique provides the possibility to resolve depth coordinates of particles shadowing each other. Particles with $(x,y)$ coordinates very close to each other, but significantly different $z$ positions (depth) will appear with maximum brightness at different refocusing parameters $\alpha$. In this case, the $B_i(\alpha)$ brightness functions should show multiple peak structures, where each of the maxima represents the $z$-coordinate of an individual particle.

\begin{acknowledgments}
This work was supported by the Hungarian Fund for Scientific Research (OTKA) through grants K77653, IN85261, K105476, NN103150.
\end{acknowledgments}


\begin{thebibliography}{15}%
\makeatletter
\providecommand \@ifxundefined [1]{%
 \@ifx{#1\undefined}
}%
\providecommand \@ifnum [1]{%
 \ifnum #1\expandafter \@firstoftwo
 \else \expandafter \@secondoftwo
 \fi
}%
\providecommand \@ifx [1]{%
 \ifx #1\expandafter \@firstoftwo
 \else \expandafter \@secondoftwo
 \fi
}%
\providecommand \natexlab [1]{#1}%
\providecommand \enquote  [1]{``#1''}%
\providecommand \bibnamefont  [1]{#1}%
\providecommand \bibfnamefont [1]{#1}%
\providecommand \citenamefont [1]{#1}%
\providecommand \href@noop [0]{\@secondoftwo}%
\providecommand \href [0]{\begingroup \@sanitize@url \@href}%
\providecommand \@href[1]{\@@startlink{#1}\@@href}%
\providecommand \@@href[1]{\endgroup#1\@@endlink}%
\providecommand \@sanitize@url [0]{\catcode `\\12\catcode `\$12\catcode
  `\&12\catcode `\#12\catcode `\^12\catcode `\_12\catcode `\%12\relax}%
\providecommand \@@startlink[1]{}%
\providecommand \@@endlink[0]{}%
\providecommand \url  [0]{\begingroup\@sanitize@url \@url }%
\providecommand \@url [1]{\endgroup\@href {#1}{\urlprefix }}%
\providecommand \urlprefix  [0]{URL }%
\providecommand \Eprint [0]{\href }%
\providecommand \doibase [0]{http://dx.doi.org/}%
\providecommand \selectlanguage [0]{\@gobble}%
\providecommand \bibinfo  [0]{\@secondoftwo}%
\providecommand \bibfield  [0]{\@secondoftwo}%
\providecommand \translation [1]{[#1]}%
\providecommand \BibitemOpen [0]{}%
\providecommand \bibitemStop [0]{}%
\providecommand \bibitemNoStop [0]{.\EOS\space}%
\providecommand \EOS [0]{\spacefactor3000\relax}%
\providecommand \BibitemShut  [1]{\csname bibitem#1\endcsname}%
\let\auto@bib@innerbib\@empty
\bibitem [{\citenamefont {Chu}\ and\ \citenamefont {I}(1994)}]{pk1}%
  \BibitemOpen
  \bibfield  {author} {\bibinfo {author} {\bibfnamefont {J.~H.}\ \bibnamefont
  {Chu}}\ and\ \bibinfo {author} {\bibfnamefont {Lin}~\bibnamefont {I}},\ }\href
  {\doibase 10.1103/PhysRevLett.72.4009} {\bibfield  {journal} {\bibinfo
  {journal} {Phys. Rev. Lett.}\ }\textbf {\bibinfo {volume} {72}},\ \bibinfo
  {pages} {4009} (\bibinfo {year} {1994})}\BibitemShut {NoStop}%
\bibitem [{\citenamefont {Thomas}\ \emph {et~al.}(1994)\citenamefont {Thomas},
  \citenamefont {Morfill}, \citenamefont {Demmel}, \citenamefont {Goree},
  \citenamefont {Feuerbacher},\ and\ \citenamefont {M\"ohlmann}}]{pk2}%
  \BibitemOpen
  \bibfield  {author} {\bibinfo {author} {\bibfnamefont {H.}~\bibnamefont
  {Thomas}}, \bibinfo {author} {\bibfnamefont {G.~E.}\ \bibnamefont {Morfill}},
  \bibinfo {author} {\bibfnamefont {V.}~\bibnamefont {Demmel}}, \bibinfo
  {author} {\bibfnamefont {J.}~\bibnamefont {Goree}}, \bibinfo {author}
  {\bibfnamefont {B.}~\bibnamefont {Feuerbacher}}, \ and\ \bibinfo {author}
  {\bibfnamefont {D.}~\bibnamefont {M\"ohlmann}},\ }\href {\doibase
  10.1103/PhysRevLett.73.652} {\bibfield  {journal} {\bibinfo  {journal} {Phys.
  Rev. Lett.}\ }\textbf {\bibinfo {volume} {73}},\ \bibinfo {pages} {652}
  (\bibinfo {year} {1994})}\BibitemShut {NoStop}%
\bibitem [{\citenamefont {Melzer}, \citenamefont {Trottenberg},\ and\
  \citenamefont {Piel}(1994)}]{pk3}%
  \BibitemOpen
  \bibfield  {author} {\bibinfo {author} {\bibfnamefont {A.}~\bibnamefont
  {Melzer}}, \bibinfo {author} {\bibfnamefont {T.}~\bibnamefont {Trottenberg}},
  \ and\ \bibinfo {author} {\bibfnamefont {A.}~\bibnamefont {Piel}},\ }\href
  {\doibase 10.1016/0375-9601(94)90144-9} {\bibfield  {journal} {\bibinfo
  {journal} {Physics Letters A}\ }\textbf {\bibinfo {volume} {191}},\ \bibinfo
  {pages} {301 } (\bibinfo {year} {1994})}\BibitemShut {NoStop}%
\bibitem [{\citenamefont {Fortov}\ \emph {et~al.}(1998)\citenamefont {Fortov},
  \citenamefont {Nefedov}, \citenamefont {Vaulina}, \citenamefont {Lipaev},
  \citenamefont {Molotkov}, \citenamefont {Samaryan}, \citenamefont
  {Nikitskii}, \citenamefont {Ivanov}, \citenamefont {Savin},\ and\
  \citenamefont {Kalmykov}}]{Fortov}%
  \BibitemOpen
  \bibfield  {author} {\bibinfo {author} {\bibfnamefont {V.~E.}\ \bibnamefont
  {Fortov}}, \bibinfo {author} {\bibfnamefont {A.~P.}\ \bibnamefont {Nefedov}},
  \bibinfo {author} {\bibfnamefont {O.~S.}\ \bibnamefont {Vaulina}}, \bibinfo
  {author} {\bibfnamefont {A.~M.}\ \bibnamefont {Lipaev}}, \bibinfo {author}
  {\bibfnamefont {V.~I.}\ \bibnamefont {Molotkov}}, \bibinfo {author}
  {\bibfnamefont {A.~A.}\ \bibnamefont {Samaryan}}, \bibinfo {author}
  {\bibfnamefont {V.~P.}\ \bibnamefont {Nikitskii}}, \bibinfo {author}
  {\bibfnamefont {A.~I.}\ \bibnamefont {Ivanov}}, \bibinfo {author}
  {\bibfnamefont {S.~F.}\ \bibnamefont {Savin}}, \ and\ \bibinfo {author}
  {\bibfnamefont {A.~V.}\ \bibnamefont {Kalmykov}},\ }\href@noop {} {\bibfield
  {journal} {\bibinfo  {journal} {JETP}\ }\textbf {\bibinfo {volume} {87}},\
  \bibinfo {pages} {1087} (\bibinfo {year} {1998})}\BibitemShut {NoStop}%
\bibitem [{\citenamefont {Nefedov}\ \emph {et~al.}(2003)\citenamefont
  {Nefedov}, \citenamefont {Morfill}, \citenamefont {Fortov}, \citenamefont
  {Thomas}, \citenamefont {Rothermel}, \citenamefont {Hagl}, \citenamefont
  {Ivlev}, \citenamefont {Zuzic}, \citenamefont {Klumov},\ and\ \citenamefont
  {Lipaev}}]{Nefedov}%
  \BibitemOpen
  \bibfield  {author} {\bibinfo {author} {\bibfnamefont {A.~P.}\ \bibnamefont
  {Nefedov}}, \bibinfo {author} {\bibfnamefont {G.~E.}\ \bibnamefont
  {Morfill}}, \bibinfo {author} {\bibfnamefont {V.~E.}\ \bibnamefont {Fortov}},
  \bibinfo {author} {\bibfnamefont {H.~M.}\ \bibnamefont {Thomas}}, \bibinfo
  {author} {\bibfnamefont {H.}~\bibnamefont {Rothermel}}, \bibinfo {author}
  {\bibfnamefont {T.}~\bibnamefont {Hagl}}, \bibinfo {author} {\bibfnamefont
  {A.~V.}\ \bibnamefont {Ivlev}}, \bibinfo {author} {\bibfnamefont
  {M.}~\bibnamefont {Zuzic}}, \bibinfo {author} {\bibfnamefont {B.~A.}\
  \bibnamefont {Klumov}}, \ and\ \bibinfo {author} {\bibfnamefont {A.~M.}\
  \bibnamefont {Lipaev}},\ }\href@noop {} {\bibfield  {journal} {\bibinfo
  {journal} {New Journal Of Physics}\ }\textbf {\bibinfo {volume} {5}},\
  \bibinfo {pages} {33} (\bibinfo {year} {2003})}\BibitemShut {NoStop}%
\bibitem [{\citenamefont {Buttensch\"on}, \citenamefont {Himpel},\ and\
  \citenamefont {Melzer}(2011)}]{parabolic}%
  \BibitemOpen
  \bibfield  {author} {\bibinfo {author} {\bibfnamefont {B.}~\bibnamefont
  {Buttensch\"on}}, \bibinfo {author} {\bibfnamefont {M.}~\bibnamefont
  {Himpel}}, \ and\ \bibinfo {author} {\bibfnamefont {A.}~\bibnamefont
  {Melzer}},\ }\href@noop {} {\bibfield  {journal} {\bibinfo  {journal} {New
  Journal of Physics}\ }\textbf {\bibinfo {volume} {13}},\ \bibinfo {pages}
  {023042} (\bibinfo {year} {2011})}\BibitemShut {NoStop}%
\bibitem [{\citenamefont {Arp}\ \emph {et~al.}(2005)\citenamefont {Arp},
  \citenamefont {Block}, \citenamefont {Bonitz}, \citenamefont {Fehske},
  \citenamefont {Golubnychiy}, \citenamefont {Kosse}, \citenamefont {Ludwig},
  \citenamefont {Melzer},\ and\ \citenamefont {Piel}}]{Cball}%
  \BibitemOpen
  \bibfield  {author} {\bibinfo {author} {\bibfnamefont {O.}~\bibnamefont
  {Arp}}, \bibinfo {author} {\bibfnamefont {D.}~\bibnamefont {Block}}, \bibinfo
  {author} {\bibfnamefont {M.}~\bibnamefont {Bonitz}}, \bibinfo {author}
  {\bibfnamefont {H.}~\bibnamefont {Fehske}}, \bibinfo {author} {\bibfnamefont
  {V.}~\bibnamefont {Golubnychiy}}, \bibinfo {author} {\bibfnamefont
  {S.}~\bibnamefont {Kosse}}, \bibinfo {author} {\bibfnamefont
  {P.}~\bibnamefont {Ludwig}}, \bibinfo {author} {\bibfnamefont
  {A.}~\bibnamefont {Melzer}}, \ and\ \bibinfo {author} {\bibfnamefont
  {A.}~\bibnamefont {Piel}},\ }\href@noop {} {\bibfield  {journal} {\bibinfo
  {journal} {J. Phys.: Conf. Ser.}\ }\textbf {\bibinfo {volume} {11}},\
  \bibinfo {pages} {234} (\bibinfo {year} {2005})}\BibitemShut {NoStop}%
\bibitem [{\citenamefont {Kong}\ \emph {et~al.}(2011)\citenamefont {Kong},
  \citenamefont {Hyde}, \citenamefont {Matthews}, \citenamefont {Qiao},
  \citenamefont {Zhang},\ and\ \citenamefont {Douglass}}]{WacoBox}%
  \BibitemOpen
  \bibfield  {author} {\bibinfo {author} {\bibfnamefont {J.}~\bibnamefont
  {Kong}}, \bibinfo {author} {\bibfnamefont {T.~W.}\ \bibnamefont {Hyde}},
  \bibinfo {author} {\bibfnamefont {L.}~\bibnamefont {Matthews}}, \bibinfo
  {author} {\bibfnamefont {K.}~\bibnamefont {Qiao}}, \bibinfo {author}
  {\bibfnamefont {Z.}~\bibnamefont {Zhang}}, \ and\ \bibinfo {author}
  {\bibfnamefont {A.}~\bibnamefont {Douglass}},\ }\href@noop {} {\bibfield
  {journal} {\bibinfo  {journal} {Phys. Rev. E}\ }\textbf {\bibinfo {volume}
  {84}},\ \bibinfo {pages} {016411} (\bibinfo {year} {2011})}\BibitemShut
  {NoStop}%
\bibitem [{\citenamefont {Samsonov}\ \emph {et~al.}(2008)\citenamefont
  {Samsonov}, \citenamefont {Elsaesser}, \citenamefont {Edwards}, \citenamefont
  {Thomas},\ and\ \citenamefont {Morfill}}]{slicing}%
  \BibitemOpen
  \bibfield  {author} {\bibinfo {author} {\bibfnamefont {D.}~\bibnamefont
  {Samsonov}}, \bibinfo {author} {\bibfnamefont {A.}~\bibnamefont {Elsaesser}},
  \bibinfo {author} {\bibfnamefont {A.}~\bibnamefont {Edwards}}, \bibinfo
  {author} {\bibfnamefont {H.~M.}\ \bibnamefont {Thomas}}, \ and\ \bibinfo
  {author} {\bibfnamefont {G.~E.}\ \bibnamefont {Morfill}},\ }\href@noop {}
  {\bibfield  {journal} {\bibinfo  {journal} {Rev. Sci. Instrum.}\ }\textbf
  {\bibinfo {volume} {79}},\ \bibinfo {pages} {035102} (\bibinfo {year}
  {2008})}\BibitemShut {NoStop}%
\bibitem [{\citenamefont {K\"{a}ding}\ and\ \citenamefont
  {Melzer}(2006)}]{stereo}%
  \BibitemOpen
  \bibfield  {author} {\bibinfo {author} {\bibfnamefont {S.}~\bibnamefont
  {K\"{a}ding}}\ and\ \bibinfo {author} {\bibfnamefont {A.}~\bibnamefont
  {Melzer}},\ }\href {\doibase 10.1063/1.2354149} {\bibfield  {journal}
  {\bibinfo  {journal} {Phys. Plasmas}\ }\textbf {\bibinfo {volume} {13}},\
  \bibinfo {eid} {090701} (\bibinfo {year} {2006})}\BibitemShut {NoStop}%
\bibitem [{\citenamefont {Annaratone}\ \emph {et~al.}(2004)\citenamefont
  {Annaratone}, \citenamefont {Antonova}, \citenamefont {Goldbeck},
  \citenamefont {Thomas},\ and\ \citenamefont {Morfill}}]{gradient}%
  \BibitemOpen
  \bibfield  {author} {\bibinfo {author} {\bibfnamefont {B.~M.}\ \bibnamefont
  {Annaratone}}, \bibinfo {author} {\bibfnamefont {T.}~\bibnamefont
  {Antonova}}, \bibinfo {author} {\bibfnamefont {D.~D.}\ \bibnamefont
  {Goldbeck}}, \bibinfo {author} {\bibfnamefont {H.~M.}\ \bibnamefont
  {Thomas}}, \ and\ \bibinfo {author} {\bibfnamefont {G.~E.}\ \bibnamefont
  {Morfill}},\ }\href@noop {} {\bibfield  {journal} {\bibinfo  {journal}
  {Plasma Phys. Control. Fusion}\ }\textbf {\bibinfo {volume} {46}},\ \bibinfo
  {pages} {B495} (\bibinfo {year} {2004})}\BibitemShut {NoStop}%
\bibitem [{\citenamefont {Kroll}\ \emph {et~al.}(2008)\citenamefont {Kroll},
  \citenamefont {Harms}, \citenamefont {Block},\ and\ \citenamefont
  {Piel}}]{holography}%
  \BibitemOpen
  \bibfield  {author} {\bibinfo {author} {\bibfnamefont {M.}~\bibnamefont
  {Kroll}}, \bibinfo {author} {\bibfnamefont {S.}~\bibnamefont {Harms}},
  \bibinfo {author} {\bibfnamefont {D.}~\bibnamefont {Block}}, \ and\ \bibinfo
  {author} {\bibfnamefont {A.}~\bibnamefont {Piel}},\ }\href {\doibase
  10.1063/1.2932109} {\bibfield  {journal} {\bibinfo  {journal} {Phys.
  Plasmas}\ }\textbf {\bibinfo {volume} {15}},\ \bibinfo {eid} {063703}
  (\bibinfo {year} {2008})}\BibitemShut {NoStop}%
\bibitem [{\citenamefont {Lippmann}(1908)}]{Lippmann}%
  \BibitemOpen
  \bibfield  {author} {\bibinfo {author} {\bibfnamefont {G.}~\bibnamefont
  {Lippmann}},\ }\href@noop {} {\bibfield  {journal} {\bibinfo  {journal}
  {Comptes Rendus de l'AcadŽmie des Sciences}\ }\textbf {\bibinfo {volume}
  {146}},\ \bibinfo {pages} {446} (\bibinfo {year} {1908})}\BibitemShut
  {NoStop}%
\bibitem [{\citenamefont {Ng}\ \emph {et~al.}(2005)\citenamefont {Ng},
  \citenamefont {Levoy}, \citenamefont {Bredif}, \citenamefont {Duval},
  \citenamefont {Horowitz},\ and\ \citenamefont {Hanrahan}}]{Ng2005}%
  \BibitemOpen
  \bibfield  {author} {\bibinfo {author} {\bibfnamefont {R.}~\bibnamefont
  {Ng}}, \bibinfo {author} {\bibfnamefont {M.}~\bibnamefont {Levoy}}, \bibinfo
  {author} {\bibfnamefont {M.}~\bibnamefont {Bredif}}, \bibinfo {author}
  {\bibfnamefont {G.}~\bibnamefont {Duval}}, \bibinfo {author} {\bibfnamefont
  {M.}~\bibnamefont {Horowitz}}, \ and\ \bibinfo {author} {\bibfnamefont
  {P.}~\bibnamefont {Hanrahan}},\ }\href@noop {} {\enquote {\bibinfo {title}
  {Light field photography with a hand-held plenoptic camera},}\ }\bibinfo
  {type} {Tech. Rep.}\ \bibinfo {number} {2005-02}\ (\bibinfo  {institution}
  {Stanford CTSR},\ \bibinfo {year} {2005})\BibitemShut {NoStop}%
\bibitem [{\citenamefont {Feng}, \citenamefont {Goree},\ and\ \citenamefont
  {Liu}(2007)}]{moment}%
  \BibitemOpen
  \bibfield  {author} {\bibinfo {author} {\bibfnamefont {Y.}~\bibnamefont
  {Feng}}, \bibinfo {author} {\bibfnamefont {J.}~\bibnamefont {Goree}}, \ and\
  \bibinfo {author} {\bibfnamefont {B.}~\bibnamefont {Liu}},\ }\href@noop {}
  {\bibfield  {journal} {\bibinfo  {journal} {Rev. Sci. Instrum.}\ }\textbf
  {\bibinfo {volume} {78}},\ \bibinfo {eid} {053704} (\bibinfo {year}
  {2007})}\BibitemShut {NoStop}%
\end{thebibliography}

\providecommand{\noopsort}[1]{}\providecommand{\singleletter}[1]{#1}%

\end{document}